\documentclass[11pt]{article}
\usepackage[a4paper,margin=28mm]{geometry}
\usepackage[T1]{fontenc}
\usepackage{lmodern}
\usepackage{amsmath,amssymb,amsthm,mathtools}
\usepackage{microtype}
\usepackage{enumitem}
\usepackage[hidelinks]{hyperref}
\usepackage{url}
\usepackage{authblk}
\setlist{topsep=5pt,itemsep=4pt}
\newtheorem{theorem}{Theorem}[section]
\newtheorem{lemma}[theorem]{Lemma}
\newtheorem{corollary}[theorem]{Corollary}

\theoremstyle{definition}
\newtheorem{definition}[theorem]{Definition}
\theoremstyle{remark}
\newtheorem{remark}[theorem]{Remark}
\newcommand{\eps}{\varepsilon}
\newcommand{\OPT}{\operatorname{OPT}}
\newcommand{\E}{\mathbb E}
\newcommand{\Prb}{\mathbb P}

\newcommand{\calE}{\mathcal E}
\newcommand{\sym}{\mathbin{\triangle}}
\newcommand{\inte}{\operatorname{int}}

\title{Single-Vertex Metropolis on Minor-Free Graphs:\\A General EPTAS}
\author[1]{Daniel Skachkov\thanks{Corresponding author.
  Email: \texttt{skach370@gmail.com}}}
\author[1]{Yuriy Dorn}
\affil[1]{MSU Institute for Artificial Intelligence, Moscow, Russia}
\date{}

\begin{document}
\maketitle

\begin{abstract}
We consider a class of combinatorial optimization problems where the goal is to find a largest induced subgraph with a prescribed property. We assume that the property is preserved by taking induced subgraphs and disjoint unions, holds for all edgeless graphs, and can be tested in polynomial time. We study the single-vertex Metropolis process for this class of problems.  We show that on graphs excluding an arbitrary fixed minor $H$, this process is a randomized efficient polynomial-time approximation scheme. Starting from any feasible solution on a graph with $n$ vertices, it finds a $(1-\eps)$-approximation
within $\exp(O_H(\eps^{-2}))\,n(n+\log(1/\delta))$ steps with
probability at least $1-\delta$.
This includes maximum independent set, maximum induced forest,
maximum induced bipartite subgraph, and maximum induced subgraph of
degree at most one. The proof uses improving exchanges on one fixed division of the input graph into regions. A recursive separator construction performs each exchange inside a region of size $r$ in at most $2r$ moves, with temporary loss $O_H(\sqrt r)$. We count move sequences through each transition and bound the
probability of avoiding all near-optimal states. The general theorem bounds the number of Metropolis steps in terms
of the temporary loss along improvement paths. We extend the result to bounded weight imbalance and give a weighted-star obstruction for unrestricted weights.
\end{abstract}


\section{Introduction}\label{sec:intro}

Many graph problems ask us to keep as many vertices as possible while
making the graph induced by those vertices satisfy some condition.
For an independent set, the condition is that no edge remains. Other
examples ask that the remaining graph have no cycle, be bipartite, or
have maximum degree at most one.

The Metropolis process maintains one feasible set, chooses a vertex uniformly at random, accepts any
feasible addition, and accepts a deletion with a prescribed probability. It returns the largest set visited. 
We prove that this process gives an efficient polynomial-time
approximation scheme on every graph class excluding a fixed minor.

A graph minor is obtained by deleting vertices or edges and contracting
edges. Contracting an edge merges its endpoints; loops and repeated
edges are then removed. A graph is \emph{$H$-minor-free} if these
operations cannot produce $H$. Throughout the main result, $H$ is fixed.
Planar graphs are included, since every planar graph excludes $K_5$.

We express the approximation guarantee through the temporary loss
needed to carry out a local improvement. This separates the probabilistic
argument from the structure of the optimization problem.

\subsection{Problem setting}\label{sec:problem}

All graphs are finite and simple. Let $\Pi$ be a fixed graph property,
and define
\[
\mathcal F_\Pi(G)=\{U\subseteq V(G):G[U]\in\Pi\},\qquad
\OPT_\Pi(G)=\max_{U\in\mathcal F_\Pi(G)}|U|.
\]
Here $G[U]$ is the induced subgraph. We call $U$ \emph{feasible} if
$G[U]\in\Pi$. We impose four conditions on $\Pi$:
\begin{enumerate}
\item \emph{Induced-subgraph closure.} Every subset of a feasible set is feasible. This condition is also called heredity.
\item \emph{Disjoint-union closure.} Combining graphs with the property,
without adding edges between them, preserves the property.
\item \emph{Edgeless graphs are feasible.} Every graph with no edges,
including the empty graph, has the property.
\item \emph{Polynomial recognition.} Whether a given graph has the
property can be decided in polynomial time.
\end{enumerate}
The first two conditions let us combine the unchanged part of a solution
with a replacement that has no edge to it. The third ensures that the
optimum contains a constant fraction of the input vertices on a fixed
minor-free class. The fourth concerns implementation.

The properties of being edgeless, being a forest, being bipartite, and having maximum degree at most one all satisfy these conditions. This fact will yield explicit corollaries in Section~\ref{sec:applications}.

\subsection{Algorithm and main result}

Let $n=|V(G)|$. If $n=0$, return the empty set. Otherwise,
start with any feasible set $U_0$, possibly empty.
For $t=0,\ldots,M-1$, make one \emph{proposal} as follows:
\begin{enumerate}
\item Choose a vertex $v$ uniformly from $V(G)$, independently of the
previous proposals.
\item If $v\notin U_t$, add it if $U_t\cup\{v\}$ is feasible;
otherwise leave the set unchanged.
\item If $v\in U_t$, remove it with probability $\lambda_t^{-1}$;
otherwise leave the set unchanged.
\end{enumerate}
Write $U_{t+1}$ for the resulting set, and return a largest set among
$U_0,\ldots,U_M$. The parameters $\lambda_t\ge1$ are the \emph{activities}. Larger activities make deletions less likely. We use constant activity and write $\lambda=\lambda_t$.

A \emph{path of single-vertex moves} from a feasible set $I$ to a feasible set $J$ is a finite sequence $U_0,U_1,\ldots,U_k$ of feasible sets, also called \emph{states}, where $U_0=I$, $U_k=J$, and each $U_i$ is obtained from $U_{i-1}$ by adding or deleting exactly one vertex. Its \emph{length} is $k$,
the number of moves.

We say that the path \emph{changes only vertices in} $R\subseteq V(G)$ if
\[
U_i\setminus R=I\setminus R
\qquad\text{for every }i=0,\ldots,k.
\]
Thus vertices outside $R$ keep their membership throughout the path,
while vertices inside $R$ may be removed and added. The sets in the sequence may repeat. The path is called \emph{simple} if all its states
are distinct.

\begin{definition}\label{def:loss}
A nondecreasing computable function
$h:\mathbb N_0\to[0,\infty)$ is a \emph{regional temporary-loss bound}
for $\Pi$ on a graph class if the following holds for every integer
$s\ge1$. Let $R$ be a vertex set with $|R|\le s$, and let $I,J$ be
feasible sets such that
\[
|J|>|I|, \qquad I\setminus R=J\setminus R,
\qquad
\text{there is no edge between }J\cap R\text{ and }I\setminus R.
\]
Then there is a path from $I$ to $J$ of at most $2s$ feasible
single-vertex moves, changing only vertices in $R$, such that each
state $L$ on the path satisfies
\[
|L|\ge |I|-h(s).
\]
The bound must hold uniformly over all input graphs, regions, and
endpoint pairs; $h$ depends only on $s$, the property $\Pi$, and the graph class.
\end{definition}

The set $J$ is a better solution than $I$, and the two sets agree
outside $R$. The definition guarantees that we can transform $I$ into $J$ by
feasible single-vertex moves within $R$, without a large temporary decrease in the size of the intermediate solutions.

A trivial regional temporary-loss bound is $h(s)=s$: first delete all vertices in $I\setminus J$, then add all vertices in $J\setminus I$. Every intermediate set is a subset of either $I$ or $J$, so it remains feasible. Since $|I\setminus J|\le |I\sym J|\le s$, the size never drops below $|I|-s$.

The following theorem relates the proposal bound to the temporary loss.

\begin{theorem}\label{thm:main}
For every fixed graph $H$, there are positive constants $R_H,A_H,C_H$
with the following property. Let $\Pi$ satisfy the four conditions in
Section~\ref{sec:problem}, and let $h$ be a regional temporary-loss bound for
$\Pi$ on $H$-minor-free graphs. For $\eps,\delta\in(0,1)$, put
\begin{equation}\label{eq:r}
r=\lceil R_H\eps^{-2}\rceil,\qquad
\lambda=\exp(A_H/\eps).
\end{equation}
On every $H$-minor-free graph on $n$ vertices, from every feasible initial set,
the Metropolis process with constant activity $\lambda$ satisfies
\begin{equation}\label{eq:success}
\Prb\!\left\{\max_{0\le t\le M}|U_t|
       \ge(1-\eps)\OPT_\Pi(G)\right\}\ge1-\delta
\end{equation}
whenever the integer proposal budget satisfies
\begin{equation}\label{eq:runtime}
M\ge
\exp\!\left(C_H\left[r+\frac{h(r)}{\eps}\right]\right)
n\bigl(n+\log(1/\delta)\bigr).
\end{equation}
\end{theorem}

Lemma~\ref{lem:regional-conversion} improves the trivial bound
$h(s)=s$ to $h(s)=O_H(\sqrt{s})$ for every property considered here. Theorem~\ref{thm:main} therefore yields
\begin{equation}\label{eq:headline}
M\ge\exp(O_H(\eps^{-2}))\,n\bigl(n+\log(1/\delta)\bigr)
\end{equation}
throughout the whole class of problems.

For fixed $H$ and $\Pi$, polynomial recognition turns this bound into
a randomized EPTAS: for fixed failure probability the running time has the form $f_H(\eps)n^{O_\Pi(1)}$, whose polynomial degree does not depend on $\eps$.

\subsection{Proof outline}\label{sec:idea}

Fix an optimal set and divide the input graph into small regions with
few shared boundary vertices. Use the same division for every state
of the Metropolis process. A solution far from optimal must be missing
many vertices that the optimum keeps. The boundaries account for only
a small part of this deficit, so many regions offer a useful improvement:
remove the current solution's vertices inside the region and insert
the optimal vertices from its interior. There are no edges from the
inserted vertices to the part left outside the region, so the closure
assumptions ensure that the replacement is feasible.

Each improvement can be carried out by single-vertex moves.
The temporary-loss bound limits how far the solution size can fall
along the path. The random process is unlikely to follow any one
prescribed path exactly. We therefore analyze all useful regional
improvements together and show that the process is unlikely to avoid
a near-optimal solution for too long.

\subsection{Contributions}\label{sec:contribution}

Our main contribution is a general randomized EPTAS guarantee for
the standard single-vertex Metropolis process on graphs excluding
a fixed minor. The result applies to every graph property satisfying
the four conditions in Section~\ref{sec:problem}. In particular, it
covers Maximum Independent Set, Maximum Induced Forest, Maximum
Induced Bipartite Subgraph, and Maximum Induced Subgraph of Degree
at Most One. For all these problems, the process finds a
$(1-\eps)$-approximation with probability at least $1-\delta$,
from every feasible initial set, within
\[
\exp(O_H(\eps^{-2}))\,n\bigl(n+\log(1/\delta)\bigr)
\]
proposals. The polynomial exponent is independent of the accuracy,
and polynomial-time feasibility testing gives a randomized EPTAS.

A central consequence resolves an explicit open question of
Chen, Mikulincer, Reichman, and Wein~\cite[Section~7]{CMRW}.
They ask what approximation ratio the Metropolis process can achieve
in polynomial time for Maximum Independent Set on planar graphs.
Corollary~\ref{cor:independent} answers this question with an EPTAS
for all planar graphs and, more generally, for every class excluding
a fixed minor. Their positive approximation result for
forests~\cite[Theorem~4]{CMRW} thus extends to a substantially broader
class, with an explicit dependence on both accuracy and failure
probability.

We also extend the argument to nonnegative vertex weights.
If $W$ is the ratio of the maximum vertex weight to the average
weight, Corollary~\ref{cor:weighted} gives a proposal bound of
\[
\exp(O_H((W/\eps)^2))\,n\bigl(n+\log(1/\delta)\bigr).
\]
For bounded $W$, this retains the dependence on accuracy of the
unweighted result. Theorem~\ref{thm:star} complements this guarantee
with an obstruction on weighted stars: for unrestricted weights, a
single run from the empty set has no uniform constant-factor guarantee
in polynomial time, even when the activity is chosen adaptively.

\subsection{Related work}\label{sec:related}

\paragraph{Approximation on minor-free graphs.}
Separator-based approximation goes back to Lipton and Tarjan~\cite{LT}.
Baker~\cite{Baker} developed layering-based approximation schemes for
planar graphs, and Alon, Seymour, and Thomas~\cite{AST} proved the
separator theorem for graphs excluding a fixed minor. Approximation
frameworks for minor-free graphs include those of Grohe~\cite{Grohe},
Demaine, Hajiaghayi, and Kawarabayashi~\cite{DHK}, and
Tazari~\cite{Tazari}. In particular, the maximum induced-subgraph
formulation for hereditary properties is treated in
\cite[Theorem 3.7]{DHK}, under its bounded-treewidth solvability
condition.

For the exact assumptions in this paper, a deterministic EPTAS also
follows directly from separators and exhaustive search on small pieces.
Delete the shared boundary of a division with region size
$O_H(\eps^{-2})$, solve each remaining interior exactly by enumerating
its subsets, and take the union. The interiors have no edges between
them. Only $O_H(n/\sqrt r)$ vertices are lost from the optimum, which
is at least $a_Hn$ for a constant $a_H>0$. Polynomial recognition makes
the enumeration an EPTAS. Our result establishes an EPTAS guarantee
for the single-vertex Metropolis process itself.

Cabello and Gajser~\cite{CG} give closely related local-search PTASs
for independent set, vertex cover, and dominating set on minor-free
graphs. Their independent-set algorithm searches for improvements
involving $O_H(\eps^{-2})$ vertices and takes
$n^{O_H(\eps^{-2})}$ time. We use their division in the analysis,
but the process itself proposes only one vertex at a time. Its
polynomial exponent is independent of the accuracy.

\paragraph{Metropolis and simulated annealing.}
The Metropolis rule originated in statistical mechanics~\cite{MRRTT};
simulated annealing developed it into an optimization method~\cite{KGV}.
Hajek's cooling criterion~\cite{Hajek} concerns asymptotic convergence.
Finite-time bounds and approximation guarantees also have a substantial
history: Sasaki and Hajek~\cite{SH} studied matching, and Desai and
Rao~\cite{DR} developed finite-time estimates for slowly cooled chains.
Doerr, Rajabi, and Witt~\cite{DRW} proved a simulated-annealing PTAS for
minimum spanning tree. Lower bounds include Jerrum's clique
example~\cite{Jerrum} and the planted-clique results of Chen, Mossel,
and Zadik~\cite{CMZ}.

For independent sets, Chen, Mikulincer, Reichman, and Wein~\cite{CMRW}
prove lower bounds for Metropolis and simulated annealing, including
exact optimization on trees. They also obtain a positive approximation
result for forests and discuss bounded treewidth, using mixing results
such as those of Eppstein and Frishberg~\cite{EF}. They explicitly ask
what approximation guarantee Metropolis can achieve in polynomial time
on planar graphs. Corollary~\ref{cor:independent} gives an EPTAS and
extends it to all fixed excluded minors.

Wild et al.~\cite[Section~V.D]{Wild} analyze slow mixing for weighted
independent sets on subdivided stars. Theorem~\ref{thm:star} gives
a finite-time upper bound on the probability of reaching the optimum
on a star from the empty set, uniformly over adaptive activity choices.

\paragraph{Optimization before equilibrium.}
Liu, Mohanty, Raghavendra, Rajaraman, and Wu~\cite{LMRRW} use locally
stationary distributions to analyze useful behavior without rapid
mixing; their independent-set application concerns triangle-free
bounded-degree graphs. Davies and Tan~\cite{DT} study independent sets
through local occupancy and short update sequences. Our guarantee applies to graphs excluding a fixed minor and gives
a $(1-\eps)$-approximation to the optimum.

The squared-difference argument below is a form of path comparison,
as developed by Diaconis and Stroock~\cite{DS}, Sinclair~\cite{Sinclair},
and Diaconis and Saloff-Coste~\cite{DSC}. We prove the version needed
for hitting a target set at constant activity. We combine it with
a fixed collection of regional improvements, bounds on the number of
paths ending at a state or using a transition, and a temporary-loss bound.

\paragraph{Independent-set reconfiguration.}
De Berg, Jansen, and Mukherjee~\cite{BJM} study the minimum intermediate
size when transforming one independent set into another. Hoang,
Ohsaka, Saito, and Tamura~\cite{HOST} give approximation schemes for
this reconfiguration objective on bounded-treewidth and minor-free
graphs. Separator-based control of intermediate sizes is therefore
established. Here the paths must also stay within the vertices of one
exchange, have bounded length, and support the regional counting
argument. We give a direct proof of these particular properties.

\section{A hitting-time bound for Metropolis}\label{sec:paths}

We first prove a hitting-time bound for a Metropolis process on any
finite state space. The graph construction enters in Section~\ref{sec:graphs}.

\subsection{Reversibility and energy}

Let $\Omega$ be a finite state space, that is, a finite set of feasible
solutions. Let $F:\Omega\to[0,\infty)$ assign an objective value to each
state. Larger values are better.

For distinct states $x,y$, let $q(x,y)$ be their proposal probability.
We assume that $q(x,y)=q(y,x)$ and
$\sum_{y\ne x}q(x,y)\le1$ for every $x\in\Omega$. The remaining probability is assigned to
proposing no change. For the single-vertex process, $q(I,J)=1/n$
when $I$ and $J$ differ by one vertex and both are feasible, and it is zero for other distinct states.

With activity $\lambda\ge1$, the probability of actually moving from
$x$ to a distinct state $y$ is
\begin{equation}\label{eq:metropolis-general}
P_\lambda(x,y)
=q(x,y)\min\{1,\lambda^{F(y)-F(x)}\}.
\end{equation}
Set $P_\lambda(x,x)=1-\sum_{y\ne x}P_\lambda(x,y)$.
The matrix $P_\lambda$ records all one-step transition probabilities. Each of its rows sums to one.

Assign to each state the positive number
\[
\omega_\lambda(x)=\lambda^{F(x)}.
\]
We call this its \emph{state weight}.
The state weights satisfy
\begin{equation}\label{eq:balance}
\omega_\lambda(x)P_\lambda(x,y)
=\omega_\lambda(y)P_\lambda(y,x)
=q(x,y)\lambda^{\min\{F(x),F(y)\}}.
\end{equation}

Identity~\eqref{eq:balance} is called \emph{reversibility} or detailed
balance. Dividing the state weights by their sum would give a stationary
probability distribution: if the process started with that distribution,
one step would leave it unchanged.

For an allowed unordered transition edge $e=\{u,v\}$, define
\begin{equation}\label{eq:conductance}
Q_\lambda(e)=\omega_\lambda(u)P_\lambda(u,v)
=q(u,v)\lambda^{\min\{F(u),F(v)\}}.
\end{equation}
Reversibility makes this independent of the orientation of the edge.
The number $Q_\lambda(e)$ is often called its conductance. In the proof
it is simply the positive coefficient assigned to that transition.

For any real-valued function $f$ on the states, set
\begin{equation}\label{eq:energy-definition}
\calE_\lambda(f,f)
=\sum_{e=\{u,v\}}Q_\lambda(e)(f(u)-f(v))^2.
\end{equation}
The sum counts each allowed unordered edge once and excludes loops.
This is the \emph{Dirichlet form}. It measures changes of $f$ across
possible moves, with the transition coefficients included.

\subsection{Improving paths}

Let $\mathcal A\subseteq\Omega$ be the set of target states, and let
$\mathcal B=\Omega\setminus\mathcal A$ be the other states.
For each $x\in\mathcal B$, fix a finite collection of
\emph{auxiliary paths}, indexed by $j$. Path $\gamma_{xj}$ starts
at $x$, ends at a state $y_{xj}$, and uses only allowed transitions.
Each path is simple: no state is repeated. Associate coefficients $p_{xj}\ge0$ with these paths, with
$\sum_jp_{xj}=1$ for each starting state $x$.

Two quantities will enter the proof. The first is
\begin{equation}\label{eq:a-definition}
a_\lambda=\max_{y\in\Omega}
\sum_{\substack{x\in\mathcal B,\,j\\y_{xj}=y}}
p_{xj}\frac{\omega_\lambda(x)}{\omega_\lambda(y)}.
\end{equation}

The second quantity is
\begin{equation}\label{eq:b-definition}
b_\lambda=\max_e\frac{1}{Q_\lambda(e)}
\sum_{\substack{x\in\mathcal B,\,j\\e\in\gamma_{xj}}}
p_{xj}\omega_\lambda(x)|\gamma_{xj}|.
\end{equation}
Here $|\gamma_{xj}|$ is the length of the path.
The quantity $a_\lambda$ measures how much weighted mass can be sent
to one endpoint. The quantity $b_\lambda$ measures how much must pass
through one transition, relative to its conductance. Small values prevent
many improvements from being forced through the same rare move.

\begin{lemma}\label{lem:flow}
If $a_\lambda\le1/4$, then every real function $f$ that is zero on
$\mathcal A$ satisfies
\begin{equation}\label{eq:dirichlet}
\sum_{x\in\Omega}\omega_\lambda(x)f(x)^2
\le4b_\lambda\calE_\lambda(f,f).
\end{equation}
\end{lemma}

\begin{proof}
Write
\[
S_f=\sum_{x\in\Omega}\omega_\lambda(x)f(x)^2
=\sum_{x\in\mathcal B}\omega_\lambda(x)f(x)^2.
\]
The second equality uses $f=0$ on $\mathcal A$.
We compare the value at each starting state with the value at the end of each of its paths.

For any real numbers $u,v$, the inequality
$u^2\le2(u-v)^2+2v^2$ follows by writing $u=(u-v)+v$.
Apply it to $u=f(x)$ and $v=f(y_{xj})$, multiply by
$\omega_\lambda(x)p_{xj}$, and sum over $x\in\mathcal B$ and $j$.
Since the coefficients for each $x$ sum to one, the result is
\begin{align}
S_f\le{}&2\sum_{x\in\mathcal B,j}
 p_{xj}\omega_\lambda(x)(f(x)-f(y_{xj}))^2\notag\\
&+2\sum_{x\in\mathcal B,j}
 p_{xj}\omega_\lambda(x)f(y_{xj})^2.\label{eq:split-function}
\end{align}
We bound these two sums separately.

Group the second sum by its final state $y$. Before the factor two, it is
\begin{align*}
\sum_{y\in\Omega} f(y)^2
 \sum_{\substack{x\in\mathcal B,j\\y_{xj}=y}}
 p_{xj}\omega_\lambda(x)
&=\sum_{y\in\Omega}\omega_\lambda(y)f(y)^2
 \sum_{\substack{x\in\mathcal B,j\\y_{xj}=y}}
 p_{xj}\frac{\omega_\lambda(x)}{\omega_\lambda(y)}\\
&\le a_\lambda S_f.
\end{align*}

For the first sum, write a path as $z_0=x,z_1,\ldots,z_L=y_{xj}$. Its endpoint difference is the sum of its consecutive differences:
\[
f(x)-f(y_{xj})=\sum_{i=0}^{L-1}(f(z_i)-f(z_{i+1})).
\]
The Cauchy--Schwarz inequality, in the form
$(a_1+\cdots+a_L)^2\le L(a_1^2+\cdots+a_L^2)$, gives
\[
(f(x)-f(y_{xj}))^2
\le |\gamma_{xj}|
\sum_{\{u,v\}\in\gamma_{xj}}(f(u)-f(v))^2.
\]
Multiply by $p_{xj}\omega_\lambda(x)$ and sum. All terms are
nonnegative, so we can group them by the edge on which the difference
is taken. By~\eqref{eq:b-definition}, the result is at most
\[
\sum_{e=\{u,v\}}b_\lambda Q_\lambda(e)(f(u)-f(v))^2
=b_\lambda\calE_\lambda(f,f).
\]

Substituting these bounds into~\eqref{eq:split-function} yields
\[
S_f\le2b_\lambda\calE_\lambda(f,f)+2a_\lambda S_f
\le2b_\lambda\calE_\lambda(f,f)+S_f/2.
\]
Rearranging proves the claim.
\end{proof}

\begin{corollary}\label{cor:counts}
Let $g>0$, $h\ge0$, $\ell\ge1$, and $B>0$.
Suppose that the auxiliary paths satisfy the following conditions.
\begin{enumerate}
\item Every path has length at most $\ell$.
\item A path starting at $x$ ends at a state of objective value at least $F(x)+g$.
\item Every state on a path starting at $x$ has objective value at least $F(x)-h$.
\item The sum of the coefficients $p_{xj}$ of paths ending at any
fixed state is at most $B$. The sum of these coefficients over paths
using any fixed transition edge is also at most $B$.
\end{enumerate}
If $q(u,v)\ge1/N$ for every allowed transition, then
\[
a_\lambda\le B\lambda^{-g},\qquad
b_\lambda\le N\ell B\lambda^h.
\]
In particular, if $\lambda^g\ge4B$, then
\begin{equation}\label{eq:kappa-form}
\sum_x\omega_\lambda(x)f(x)^2
\le\kappa^{-1}\calE_\lambda(f,f),\qquad
\kappa=\frac{1}{4N\ell B\lambda^h},
\end{equation}
for every $f$ equal to zero on $\mathcal A$.
\end{corollary}

\begin{proof}
For a path ending at $y=y_{xj}$, its gain implies
\[
\frac{\omega_\lambda(x)}{\omega_\lambda(y)}
=\lambda^{F(x)-F(y)}\le\lambda^{-g}.
\]
The coefficient sum at $y$ is at most $B$, which proves the bound on
$a_\lambda$.

For an edge $e=\{u,v\}$ on a path from $x$, both $F(u)$ and $F(v)$
are at least $F(x)-h$. By~\eqref{eq:conductance},
\[
Q_\lambda(e)\ge\frac1N\lambda^{F(x)-h},\qquad
\frac{\omega_\lambda(x)}{Q_\lambda(e)}\le N\lambda^h.
\]
Multiply this by the path length bound $\ell$ and the coefficient
sum bound $B$ to obtain the bound on $b_\lambda$.
The condition $\lambda^g\ge4B$ makes $a_\lambda\le1/4$, so
Lemma~\ref{lem:flow} proves~\eqref{eq:kappa-form}.
\end{proof}

\subsection{Hitting the target}

Let $X_t$ be the random state after $t$ proposals. Its first visit to
the target is
\[
\tau_{\mathcal A}=\inf\{t\ge0:X_t\in\mathcal A\}.
\]

\begin{lemma}\label{lem:tail}
Let $m\ge1$ be an integer, and let the activity be constant
and equal to $\lambda$.
Suppose that some $0<\kappa\le1$ satisfies
\begin{equation}\label{eq:common-dirichlet}
\sum_x\omega_\lambda(x)f(x)^2
\le\kappa^{-1}\calE_\lambda(f,f)
\end{equation}
for every $f$ zero on $\mathcal A$.
Then, from any $x_0\notin\mathcal A$,
\begin{equation}\label{eq:tail}
\Prb_{x_0}\{\tau_{\mathcal A}>m\}
\le
\left(\sum_{x\in\Omega}\lambda^{F(x)}\right)^{1/2}
\lambda^{-F(x_0)/2}e^{-\kappa m}.
\end{equation}
\end{lemma}

\begin{proof}
Write $\omega=\omega_\lambda$.

\paragraph{The killed chain.}
Let $K$ be the submatrix of $P_\lambda$ whose rows and columns
are indexed by $\mathcal B$. Its entries are nonnegative, but its rows
may sum to less than one. The missing part of a row is precisely the
probability of moving into $\mathcal A$. This is often called the
transition matrix of the chain killed on entering $\mathcal A$.

For functions on $\mathcal B$, define the weighted inner product and
norm by
\[
\langle f,g\rangle_\omega
=\sum_{x\in\mathcal B}\omega(x)f(x)g(x),\qquad
\|f\|_\omega^2=\langle f,f\rangle_\omega.
\]
Multiplication of a function by the matrix means
$(Kf)(x)=\sum_{y\in\mathcal B}K(x,y)f(y)$.
Extend $f$ by zero on $\mathcal A$. Expanding squares in the Dirichlet
form, using that rows of $P_\lambda$ sum to one, and using detailed
balance gives
\begin{align}
\calE_\lambda(f,f)
&=\sum_{e=\{x,y\}}Q_\lambda(e)(f(x)-f(y))^2\notag\\
&=\frac12 \sum_{e=\{x,y\}}(\omega(x)P_\lambda (x, y) + \omega(y)P_\lambda(y,x))(f(x)-f(y))^2\notag\\
&=\frac12\sum_{x,y\in\Omega}
 \omega(x)P_\lambda(x,y)(f(x)-f(y))^2\notag\\
&=\langle f,f\rangle_\omega
 -\langle f,Kf\rangle_\omega.\label{eq:dirichlet-identity}
\end{align}

Combining~\eqref{eq:common-dirichlet} and~\eqref{eq:dirichlet-identity}
therefore gives
\begin{equation}\label{eq:quadratic-contraction}
\langle f,Kf\rangle_\omega
\le(1-\kappa)\|f\|_\omega^2.
\end{equation}

\paragraph{Norm contraction.}
We need to show that applying $K$ decreases the norm of a function.
First, since every entry of $K$ is nonnegative, replacing function
values by their absolute values gives
\[
|\langle f,Kf\rangle_\omega|
\le\langle |f|,K|f|\rangle_\omega
\le(1-\kappa)\|f\|_\omega^2.
\]
The last step is~\eqref{eq:quadratic-contraction} applied to $|f|$.
Detailed balance also gives the symmetry identity
$\langle f,Kg\rangle_\omega=\langle g,Kf\rangle_\omega$,
by interchanging the two indices in the sum.
Consequently
\[
4\langle g,Kf\rangle_\omega
=\langle f+g,K(f+g)\rangle_\omega
-\langle f-g,K(f-g)\rangle_\omega.
\]
If $f$ and $g$ both have norm one, the preceding absolute bound gives
\begin{align*}
4|\langle g,Kf\rangle_\omega|
&\le(1-\kappa)
 \bigl(\|f+g\|_\omega^2+\|f-g\|_\omega^2\bigr)\\
&=4(1-\kappa).
\end{align*}
For a unit-norm $f$ with $Kf\ne0$, take
$g=Kf/\|Kf\|_\omega$. The left-hand inner product is then
$\|Kf\|_\omega$. The case $Kf=0$ is immediate, and scaling
handles every other $f$. We have proved
\begin{equation}\label{eq:function-contraction}
\|Kf\|_\omega\le(1-\kappa)\|f\|_\omega.
\end{equation}

\paragraph{The tail bound.}
Let $\mathbf 1$ be the function equal to one on $\mathcal B$.
The row sum of $K^m$ at $x$ is the probability of remaining outside
$\mathcal A$ for the first $m$ steps when starting from $x$. Thus
\[
p_m(x):=\Prb_x\{\tau_{\mathcal A}>m\}=(K^m\mathbf 1)(x).
\]
Applying~\eqref{eq:function-contraction} repeatedly gives
\[
\|p_m\|_\omega
\le(1-\kappa)^m\|\mathbf 1\|_\omega
\le e^{-\kappa m}
\left(\sum_{x\in\mathcal B}\omega(x)\right)^{1/2}.
\]
The term $\omega(x_0)p_m(x_0)^2$ is at most the sum
$\|p_m\|_\omega^2$. Divide by $\sqrt{\omega(x_0)}$ and enlarge
the sum from $\mathcal B$ to $\Omega$ to obtain~\eqref{eq:tail}.
\end{proof}

For feasible vertex sets with cardinality as the objective, the sum in this
tail bound has a simple estimate. Summing over all vertex subsets only
increases it, and expansion of a product gives
\begin{equation}\label{eq:partition-cardinality}
\sum_{I\in\mathcal F_\Pi(G)}\lambda^{|I|}
\le\sum_{U\subseteq V}\lambda^{|U|}
=\prod_{v\in V}(1+\lambda)=(1+\lambda)^n.
\end{equation}
Each factor corresponds to either excluding a vertex, with contribution
one, or including it, with contribution $\lambda$.

\section{Regional improvements}\label{sec:graphs}

We now construct the improving paths for feasible vertex sets on
minor-free graphs and bound how many paths can share a state or transition.

\subsection{Separators and divisions}

\begin{theorem}[Alon--Seymour--Thomas~\cite{AST}]\label{thm:separator}
For every fixed graph $H$, there is $c_H>0$ such that every
$H$-minor-free graph on $s\ge1$ vertices has a set of at most
$c_H\sqrt s$ vertices whose removal leaves components of size at most
$2s/3$.
\end{theorem}
Such a set is called a \emph{balanced separator}.

A fixed minor-free class has bounded degeneracy, by the excluded-minor
edge bounds~\cite{Thomason}. Thus there is a constant $\chi_H$ such
that every graph in the class can be properly colored with at most
$\chi_H$ colors. Indeed, repeatedly remove a vertex of bounded degree,
then color in reverse order. Vertices of one color form an independent
set. Put $a_H=1/\chi_H$. Since every edgeless graph has property $\Pi$,
\begin{equation}\label{eq:linear-alpha}
\OPT_\Pi(G)\ge\alpha(G)\ge a_H n.
\end{equation}
For nonnegative vertex weights $w_v$, write
$w(A)=\sum_{v\in A}w_v$. A heaviest color class similarly gives
\begin{equation}\label{eq:weighted-color}
\max_{U\in\mathcal F_\Pi(G)}\sum_{v\in U}w_v
\ge a_H\sum_{v\in V}w_v.
\end{equation}

A \emph{division} is a collection of regions $R_1,\ldots,R_K\subseteq V$
covering $V$. Define
\[
\partial R_j=\{v\in R_j:v\in R_i\text{ for some }i\ne j\},
\qquad \inte R_j=R_j\setminus\partial R_j.
\]
The boundary consists of shared vertices, while an interior vertex belongs
to exactly one region. We call a division \emph{closed} if every neighbor of an interior
vertex belongs to its region.

\begin{lemma}\label{lem:division}
There are constants $r_H,c_0,c_1>0$, depending only on $H$, such that
for every integer $r_H\le r\le n$, each $H$-minor-free
graph on $n$ vertices has a closed division satisfying
\begin{equation}\label{eq:division}
|R_j|\le r,\qquad K\le c_0n/r,\qquad
b:=\sum_j|\partial R_j|\le c_1n/\sqrt r.
\end{equation}
\end{lemma}

\begin{proof}
The size and boundary bounds and the closedness are supplied by
Cabello and Gajser~\cite[Lemma 2.1]{CG}. Their construction splits any
region larger than $r$ with a separator and puts the separator in both
children. Each child has at least one third of the parent's vertices.
Thus each final region has at least $r/3$ vertices.
If $m(v)$ is the number of regions containing $v$, then
\[
Kr/3\le\sum_j|R_j|=n+\sum_v(m(v)-1)\le n+b.
\]
The bound on $b$ gives $K\le c_0n/r$ for a sufficiently large $c_0$.
\end{proof}

\subsection{Regional conversion}

\begin{lemma}\label{lem:regional-conversion}
Let $\Pi$ be closed under induced subgraphs and disjoint unions.
Let $G$ be $H$-minor-free, let $R\subseteq V(G)$ have size $s$, and
let $I,J$ be feasible sets satisfying
\[
F:=I\setminus R=J\setminus R
\]
with no edges between $F$ and $J\cap R$.
Give the vertices nonnegative weights, with $w_v\le W$ for $v\in R$.
There is a simple path of single-vertex moves from $I$ to $J$, changing only
vertices in $R$, of length at most
\[
|I\cap R|+|J\cap R|\le2s,
\]
on which every state $L$ satisfies
\begin{equation}\label{eq:regional-loss}
w(L)\ge\min\{w(I),w(J)\}-c'_HW\sqrt s.
\end{equation}
The constant $c'_H$ depends only on $H$.
\end{lemma}

\begin{proof}
We define a recursive procedure $\operatorname{Convert}(X,Y)$,
where $X$ and $Y$ are vertex sets. Its output is a sequence of
single-vertex deletions and additions that transforms $X$ into $Y$.

A call works as follows. If $X=Y$, it terminates immediately.
Otherwise, write $Z=X\cup Y$, choose a balanced separator $S$
of $G[Z]$ from Theorem~\ref{thm:separator} such that
\[
|S|\le c_H\sqrt{|Z|},
\]
and let $C_1,\ldots,C_q$ be the components of $G[Z\setminus S]$. If $Z\setminus S=\varnothing$, set $q=1$ and
$C_1=\varnothing$. Each component has size at most $2|Z|/3<|Z|$.
Order the components in nonincreasing order of
\[
d_i=w(Y\cap C_i)-w(X\cap C_i).
\]
The call performs the following operations:
\begin{enumerate}
\item Delete all vertices of $X\cap S$.
\item For $i=1,\ldots,q$, in this order, make the recursive call
\[
\operatorname{Convert}(X\cap C_i,Y\cap C_i).
\]
\item Add all vertices of $Y\cap S$.
\end{enumerate}
In particular, vertices of $X\cap Y\cap S$ are deleted and later
restored. The union of the two arguments of the child call on $C_i$
is exactly $C_i$, so the recursion terminates.

We claim that applying the output of
$\operatorname{Convert}(I\cap R,J\cap R)$ to the full starting
state $I$ gives the required path of single-vertex moves,
after erasing any loops in the states.

\paragraph{Feasibility.}
We use induction on recursion depth, from the deepest calls towards
the root. For every call $\operatorname{Convert}(X,Y)$, we prove that
each connected component induced by an intermediate state is contained
in $X$ or in $Y$. For terminal calls, this is immediate since $X=Y$
and no move is made.

Now fix a nonterminal call $\operatorname{Convert}(X,Y)$.
Write $X_i=X\cap C_i$ and $Y_i=Y\cap C_i$. Recall how the procedure works.
First, it transforms $X$ into $\bigsqcup_i X_i$ by deleting
the vertices of $X\cap S$. Each such state is a subset of $X$.
Then, for $i=1,\ldots,q$, it calls $\operatorname{Convert}(X_i,Y_i)$.
During these calls, its state is the disjoint union of the child states.
There are no edges between different $C_i$, so the induction hypothesis
gives the required property. Finally, the procedure transforms
$\bigsqcup_i Y_i$ into $Y$ by adding the missing vertices. Every state
in this last stage is a subset of $Y$.

For the original call $\operatorname{Convert}(I\cap R,J\cap R)$, combine $F$ with all these connected components contained
in $I\cap R$. Their union is a subset of $I$ and hence is feasible.
The remaining parts form a subset of $J\cap R$ and have no edges
to that union, since there are no edges between $F$ and $J\cap R$ by assumption. Thus the full state is feasible at any moment by disjoint-union closure.

\paragraph{Length.}
The length bound follows from the fact that each vertex of
$I\cap R$ is deleted at most once, and each vertex of $J\cap R$
is added at most once.

\paragraph{Temporary loss.}
Set
\[
c'_H=\frac{c_H}{1-\sqrt{2/3}}.
\]
Consider any call $\operatorname{Convert}(X,Y)$. We prove that
the weight of each intermediate state
never falls below
\[
\min\{w(X),w(Y)\}-c'_HW\sqrt{|Z|}.
\]
We again use induction on the recursion depth, from the deepest
calls towards the root. For a terminal call, the claim is immediate
since its only state is $X=Y$.

Now fix a nonterminal call and choose one of its states of minimum
weight. Since all weights are nonnegative, we can choose this state
after all vertices of $X\cap S$ have been deleted and before any
vertices of $Y\cap S$ have been added.

The chosen state occurs during some child call
$\operatorname{Convert}(X_j,Y_j)$.
At this moment, the state is the disjoint union of $Y_i$ for $i<j$,
$X_i$ for $i>j$, and a state $Q_j$ of
$\operatorname{Convert}(X_j,Y_j)$.
By the induction hypothesis,
\[
w(Q_j)\ge
\min\{w(X_j),w(Y_j)\}-c'_HW\sqrt{|C_j|}.
\]

Since the gains $d_i$ are in nonincreasing order, their partial sums
attain their minimum at one of the two endpoints. Hence the full state
just before and just after each child call has weight at least
\[
\min\{w(X\setminus S),w(Y\setminus S)\}.
\]
In particular, this holds for the child call on $C_j$.
The inductive bound allows a loss of at most $c'_HW\sqrt{|C_j|}$
below the smaller of these two weights. Thus the chosen state has
weight at least
\[
\min\{w(X), w(Y)\} - W|S| - c'_HW\sqrt{|C_j|}.
\]

The inequalities $|S|\le c_H\sqrt{|Z|}$ and
$|C_j|\le2|Z|/3$, together with the choice of $c'_H$, complete
the induction.

For the original call, $|Z|\le s$, and adding the unchanged
exterior weight $w(F)$ gives~\eqref{eq:regional-loss}, since
\[
w(F)+\min\{w(I\cap R),w(J\cap R)\}
=\min\{w(I),w(J)\}.
\]
All moves lie in $R$, and erasing loops preserves the endpoints
and all the stated bounds.
\end{proof}

\subsection{Exchanges and counting}

Throughout this subsection, let $\Pi$ be closed under induced
subgraphs and disjoint unions. Give the vertices weights
$0\le w_v\le W$, where $W\ge1$ and $\sum_vw_v=n$, and write
$F(U)=\sum_{v\in U}w_v$.
Fix $\eps\in(0,1)$, an optimum $O$ with $F(O)\ge an$ for some
$a>0$, and a closed division satisfying
\begin{equation}\label{eq:regional-assumptions}
|R_j|\le r,\qquad K\le c_0n/r,\qquad
b:=\sum_j|\partial R_j|\le\frac{a\eps n}{2W}.
\end{equation}
The optimum and the division are the same for all states.
Taking $w_v=1$ and $W=1$ gives the cardinality case.

Call a feasible state $I$ bad if $F(I)<(1-\eps)F(O)$.
For each bad state and each region, define the replacement and its gain by
\begin{equation}\label{eq:region-endpoint}
J_j(I)=(I\setminus R_j)\cup(O\cap\inte R_j),
\qquad
d_j(I)=F(J_j(I))-F(I).
\end{equation}
We call a region useful for $I$ if its gain is at least $g$, where
\begin{equation}\label{eq:thresholds}
g=\frac{a\eps r}{4c_0},
\qquad
\mathcal J(I)=\{j:d_j(I)\ge g\}.
\end{equation}

The following lemma shows that there are many useful regions.

\begin{lemma}\label{lem:regional}
For every bad state $I$, all replacements $J_j(I)$ are feasible, and
\begin{equation}\label{eq:many}
|\mathcal J(I)|\ge\frac{a\eps n}{4Wr}.
\end{equation}
\end{lemma}

\begin{proof}
The sets $I\setminus R_j$ and $O\cap\inte R_j$ are feasible by
heredity. The division is closed, hence there is no edge between them. Their union is therefore feasible
by disjoint-union closure.

Let $S=\bigcup_j\partial R_j$, and let $m(v)$ be the number of
regions containing $v$. Summing the gains gives
\[
\sum_jd_j(I)
=F(O)-F(I)-F(O\cap S)
 -\sum_{v\in I\cap S}(m(v)-1)w_v.
\]
For each $v\in S$, the last two terms subtract at most $m(v)w_v$.
Since $\sum_{v\in S}m(v)=b$, we obtain
\begin{equation}\label{eq:gains}
\sum_jd_j(I)
\ge F(O)-F(I)-Wb
\ge\frac{a\eps n}{2}.
\end{equation}

Every gain is at most $Wr$, and gains outside $\mathcal J(I)$
are less than $g$. Hence
\[
\sum_jd_j(I)
\le Kg+Wr|\mathcal J(I)|
\le\frac{a\eps n}{4}+Wr|\mathcal J(I)|.
\]
Combining this with~\eqref{eq:gains} proves~\eqref{eq:many}.
\end{proof}

\begin{lemma}
\label{lem:regional-counts}
For each bad state $I$ and each $j\in\mathcal J(I)$, choose one
simple feasible path $\gamma_{Ij}$ from $I$ to $J_j(I)$ that
changes only vertices in $R_j$. Put $p_{Ij}=1/|\mathcal J(I)|$ and
\begin{equation}\label{eq:B}
B=\frac{4c_0W}{a\eps}\,2^r.
\end{equation}
Then, for every feasible state $J$ and every transition edge $e$,
\begin{align}
\sum_{\substack{I\text{ bad},\,j\in\mathcal J(I)\\J_j(I)=J}}p_{Ij}
&\le B,\label{eq:endpoint-count}\\
\sum_{\substack{I\text{ bad},\,j\in\mathcal J(I)\\e\in\gamma_{Ij}}}p_{Ij}
&\le B.\label{eq:edge-count}
\end{align}
\end{lemma}

\begin{proof}
The coefficients for each starting
state sum to one, and, by Lemma~\ref{lem:regional},
\[
p_{Ij}\le\frac{4Wr}{a\eps n}.
\]

Fix a region label $j$. If $J_j(I)=J$, then
$I\setminus R_j=J\setminus R_j$, so only $I\cap R_j$ is unknown.
There are at most $2^{|R_j|}\le2^r$ possibilities for $I$.

Similarly, if $\gamma_{Ij}$ uses an edge $e=\{L_1,L_2\}$, then
$I\setminus R_j=L_1\setminus R_j$, because the path changes
nothing outside $R_j$. This again leaves at most $2^r$
possibilities for $I$.

There is only one chosen path for each pair $(I,j)$.
Summing either count over the $K$ region labels therefore gives
\[
K\,2^r\,\frac{4Wr}{a\eps n}
\le\frac{4c_0W}{a\eps}\,2^r=B.
\]
\end{proof}

Paths can be made simple by erasing loops, without increasing
their length or temporary loss. The counting argument uses only
the fact that each path stays within its region.

Averaging over useful regions cancels the factor $n/r$ in the
number of region labels. This is why $B$ does not depend on $n$.

\section{Proof of the main theorem}\label{sec:proof}

\begin{proof}[Proof of Theorem~\ref{thm:main}]
The empty graph is immediate. Assume $n\ge1$. Take
$\Omega=\mathcal F_\Pi(G)$, $F(I)=|I|$, and
\[
\mathcal A=\{I\in\mathcal F_\Pi(G):|I|\ge(1-\eps)\OPT_\Pi(G)\}.
\]
If the initial state is already in $\mathcal A$, the output records
it. Otherwise apply Section~\ref{sec:paths}. Every allowed proposal
has probability $1/n$, so $N=n$ in Corollary~\ref{cor:counts}.
Choose $R_H$ large enough that
\[
r=\lceil R_H\eps^{-2}\rceil\ge r_H,
\qquad c_1/\sqrt r\le a_H\eps/2.
\]
These choices depend only on $H$.

\paragraph{\boldmath The case $n\ge r$.}
Fix a division from Lemma~\ref{lem:division} and an optimal set $O$.
Apply Lemmas~\ref{lem:regional} and~\ref{lem:regional-counts} with $w_v=1$, $W=1$, and $a=a_H$.
The boundary condition follows from the choice of $r$, and
$|O|\ge a_Hn$ follows from~\eqref{eq:linear-alpha}.
Each useful replacement agrees with $I$ outside $R_j$, and its
selected vertices inside $R_j$ lie in the interior. They have no edges
to $I\setminus R_j$. It improves cardinality, so
Definition~\ref{def:loss} supplies a path of length at most $2r$
with loss at most $h(r)$. Thus the parameters for Corollary~\ref{cor:counts} are
\begin{equation}\label{eq:main-parameters}
N=n,\quad \ell=2r,\quad
g=\frac{a_H\eps r}{4c_0},\quad
h=h(r),\quad
B=\frac{4c_0}{a_H\eps}\,2^r.
\end{equation}
The paths may remove and restore common selected vertices inside
the region. The counting argument uses only their confinement to $R_j$.

The gain condition is $\lambda^g\ge4B$. Its logarithmic form is
\[
g\log\lambda\ge\log(4B)
=r\log2+\log\!\left(\frac{16c_0}{a_H\eps}\right).
\]
The right-hand side is $O_H(r)$, since
$r\ge R_H\eps^{-2}$. The left-hand side for
$\lambda=\exp(A_H/\eps)$ equals $a_HA_Hr/(4c_0)$.
A sufficiently large $A_H$, depending only on $H$, therefore works
for every $\eps\in(0,1)$.

Corollary~\ref{cor:counts} therefore gives the constant
in~\eqref{eq:common-dirichlet} as
\begin{equation}\label{eq:gap}
\kappa=\frac{1}{8nrB\lambda^{h(r)}}
\ge\frac1n\exp\!\left(-D_H\left[r+\frac{h(r)}{\eps}\right]\right).
\end{equation}
Indeed, after removing the factor $n$ from $\kappa^{-1}$, its
logarithm is
$\log(8r)+\log B+A_Hh(r)/\eps$; the first two terms are $O_H(r)$.

\paragraph{\boldmath The case $1\le n<r$.}
For each bad set $I$, take one path directly to a fixed optimum $O$,
with coefficient one. Apply Definition~\ref{def:loss} with $R=V(G)$,
so the fixed exterior is empty. Its gain exceeds
$\eps\OPT_\Pi(G)\ge\eps a_Hn$. Its length is at most $2n$, and
its loss is at most $h(n)\le h(r)$. There are at most $2^n$ feasible
starting sets, so both coefficient sums are at most $2^n$.
Use $N=n$, $\ell=2n$, $B=2^n$, and $g=\eps a_Hn$ in
Corollary~\ref{cor:counts}. Increasing $A_H$ so that
$A_Ha_H\ge\log8$ ensures
\[
\lambda^{\eps a_Hn}=\exp(A_Ha_Hn)\ge4\cdot2^n
\quad(n\ge1).
\]
Corollary~\ref{cor:counts} now gives
\[
\kappa=\frac{1}{8n^2 2^n\lambda^{h(n)}}.
\]
After removing one factor $n$, the logarithm of its denominator is
\[
\log(8n)+n\log2+A_Hh(n)/\eps
=O_H\bigl(r+h(r)/\eps\bigr).
\]
Thus~\eqref{eq:gap}, with a larger $D_H$, holds in this case too.

\paragraph{The tail probability.}
Apply Lemma~\ref{lem:tail}, using
\eqref{eq:partition-cardinality}:
\begin{equation}\label{eq:failure-main}
\Prb\{\tau_{\mathcal A}>M\}
\le\exp\!\left(\frac n2\log(1+\lambda)-\kappa M\right).
\end{equation}
It suffices to take
\[
M\ge\kappa^{-1}
 \left(\frac n2\log(1+\lambda)+\log(1/\delta)\right).
\]
Since $\log(1+\lambda)\le\log2+A_H/\eps$, this is at most
\[
n\exp\!\left(D_H\left[r+\frac{h(r)}{\eps}\right]\right)
\left(C'_Hn/\eps+\log(1/\delta)\right).
\]
The extra factor $1/\eps$ is absorbed into $\exp(O_H(r))$.
Increasing $C_H$ gives~\eqref{eq:runtime}. A visit to $\mathcal A$
is retained in the largest recorded set, proving~\eqref{eq:success}.
\end{proof}

\begin{remark}
Each proposal requires polynomial work on the graph by the
recognition assumption on $\Pi$. The optimum, division, separators,
and auxiliary paths are used only in the analysis.

For an implementation using fair random bits, take
$\widehat\lambda=2^k$, where $k=\lceil E_H/\eps\rceil$ and $E_H$
is a sufficiently large positive integer depending only on $H$.
Then $\widehat\lambda\ge\lambda$ and
$\log\widehat\lambda=O_H(1/\eps)$, so the same proof applies after
increasing $C_H$. Accept a deletion when $k$ independent fair bits
are all zero; this gives probability exactly $\widehat\lambda^{-1}$.
Together with polynomial-time feasibility testing, these costs
preserve the EPTAS form.
\end{remark}

\section{Applications}
\label{sec:applications}

\begin{corollary}\label{cor:universal}
For every fixed graph $H$, there are constants $A_H,C_H>0$
with the following property.
Let $\Pi$ satisfy the four conditions in
Section~\ref{sec:problem}, and let $\eps,\delta\in(0,1)$.
On every $H$-minor-free graph on $n$ vertices, single-vertex Metropolis
with constant activity $\lambda=\exp(A_H/\eps)$, starting from
any feasible set, returns a $(1-\eps)$-approximation with
probability at least $1-\delta$ within
\begin{equation}\label{eq:universal-runtime}
M=\left\lceil
\exp(C_H\eps^{-2})\,n\bigl(n+\log(1/\delta)\bigr)
\right\rceil
\end{equation}
proposals.
\end{corollary}
\begin{proof}
Lemma~\ref{lem:regional-conversion} gives $h(s)=c'_H\sqrt s$.
Since $r=O_H(\eps^{-2})$, we have
$r+h(r)/\eps=O_H(\eps^{-2})$, so
Theorem~\ref{thm:main} yields~\eqref{eq:universal-runtime}.
\end{proof}

\begin{corollary}\label{cor:applications}
For every fixed graph $H$, single-vertex Metropolis provides
randomized EPTASs on $H$-minor-free graphs for:
\begin{enumerate}
\item Maximum Independent Set;
\item Maximum Induced Forest;
\item Maximum Induced Bipartite Subgraph;
\item Maximum Induced Subgraph of Degree at Most One
      (Maximum Dissociation Set).
\end{enumerate}
\end{corollary}

\begin{proof}
The graph properties defining the four listed problems satisfy
all four conditions in Section~\ref{sec:problem}.
Apply Corollary~\ref{cor:universal}.
\end{proof}

\begin{corollary}\label{cor:independent}
Single-vertex Metropolis is a randomized EPTAS for
Maximum Independent Set on planar graphs.
\end{corollary}

\begin{proof}
Apply Corollary~\ref{cor:applications} with $H=K_5$.
\end{proof}

\section{Weighted objectives}\label{sec:extensions}

\subsection{Bounded weight imbalance}

Let $w_v\ge0$, and write
\[
\OPT_{\Pi,w}(G)=\max\{w(U):G[U]\in\Pi\},
\qquad w(U)=\sum_{v\in U}w_v.
\]
If the total weight is zero, every feasible set is optimal. Otherwise
put
\[
\overline w=\frac1n\sum_{v\in V}w_v,
\qquad W=\frac{\max_v w_v}{\overline w}\ge1.
\]
Thus $W$ measures how large the heaviest vertex is relative to the
average. It is unchanged by scaling all weights.
Weighted Metropolis proposes a uniformly chosen vertex, accepts every
feasible addition, and accepts deletion of $v$ with probability
$\lambda^{-w_v/\overline w}$, where $\lambda\ge1$ remains fixed
throughout the run. It returns a heaviest visited set.

\begin{corollary}\label{cor:weighted}
For every fixed graph $H$, there are constants $A_H,C_H>0$
with the following property. Let $G$ be an $H$-minor-free graph with
nonnegative vertex weights of positive total weight, let
$0<\eps,\delta<1$, and let $\Pi$ satisfy the four conditions in
Section~\ref{sec:problem}. Set the constant activity to
\[
\lambda=\exp(A_H/\eps).
\]
From any feasible initial set, weighted Metropolis visits a set of
weight at least $(1-\eps)\OPT_{\Pi,w}(G)$ with probability at least
$1-\delta$ within
\[
M=\left\lceil\exp\!\left(C_H(W/\eps)^2\right)
n\bigl(n+\log(1/\delta)\bigr)\right\rceil
\]
proposals. For fixed $W$ and rational weights encoded in binary, a
finite-precision implementation is a randomized EPTAS.
\end{corollary}

\begin{proof}
We apply Section~\ref{sec:paths} to the normalized weighted objective
and use the weighted conversion bound. Normalize the weights by
$\widehat w_v=w_v/\overline w$, and write
$\widehat w(U)=\sum_{v\in U}\widehat w_v$. Then
\[
0\le\widehat w_v\le W,\qquad \sum_v\widehat w_v=n.
\]
A heaviest color class is feasible because $\Pi$ contains all edgeless
graphs. The bounded-coloring argument therefore gives
$\OPT_{\Pi,\widehat w}(G)\ge a_Hn$.

Lemma~\ref{lem:regional-conversion} gives, for every admissible
property, paths inside a region of size $s$ with at most $2s$ moves
and weighted loss
\[
L(s)=c'_HW\sqrt s.
\]

Choose $r=\lceil D_H(W/\eps)^2\rceil$, with $D_H$ large enough that
$r\ge r_H$ and $Wc_1/\sqrt r\le a_H\eps/2$.
If $n\ge r$, Lemma~\ref{lem:regional} applies with these normalized
weights and gives
\[
g=\frac{a_H\eps r}{4c_0},\qquad
B=\frac{4c_0W}{a_H\eps}2^r,\qquad \ell\le2r,
\]
with loss at most $L(r)$. The target inside a region is contained
in its interior, so the exterior condition of the conversion lemma holds.
Set $\lambda=\exp(A_H/\eps)$.
The condition $\lambda^g\ge4B$ follows with $A_H$ independent of
$W$: its logarithmic left side is $a_HA_Hr/(4c_0)$, whereas
\[
\log(4B)=r\log2+\log\!\left(\frac{16c_0W}{a_H\eps}\right)=O_H(r).
\]
Corollary~\ref{cor:counts} gives
$\kappa=(8nrB\lambda^{L(r)})^{-1}$. Writing $x=W/\eps\ge1$,
we have $r=O_H(x^2)$ and
\[
\log(\kappa^{-1}/n)
=O_H\!\left(r+\frac{L(r)}{\eps}\right)
=O_H(x^2).
\]

If $1\le n<r$, use one path from each bad set directly to a fixed
weighted optimum, using the conversion lemma with region $V(G)$
and empty exterior. Its gain exceeds $a_H\eps n$, its length is at
most $2n$, its loss is at most $L(n)\le L(r)$, and both counts are
at most $2^n$. Taking $A_Ha_H\ge\log8$ verifies the gain condition.
The resulting constant
$\kappa=(8n^2 2^n\lambda^{L(n)})^{-1}$ has the same bound because
$n<r$. Thus both graph-size cases are covered.

For the tail estimate, summing over all subsets gives
\[
\sum_{U:\,G[U]\in\Pi}\lambda^{\widehat w(U)}
\le\prod_{v\in V}(1+\lambda^{\widehat w_v})
\le2^n\lambda^n.
\]
Lemma~\ref{lem:tail} now needs at most
$\kappa^{-1}(O_H(n/\eps)+\log(1/\delta))$ proposals.
The extra factor $1/\eps$ is absorbed into the exponential bound.
Finally, normalization does not change approximation ratios, and
$\lambda^{-\widehat w_v}=\lambda^{-w_v/\overline w}$ is exactly the deletion
probability of the stated weighted process.
\end{proof}

\begin{remark}
For the algorithmic statement, assume that the weights and the
parameters $\eps,\delta$ are rational and encoded in binary.
Take $\widehat\lambda=2^k$, where
$k=\lceil E_H/\eps\rceil$ and $E_H$ is a sufficiently large
positive integer depending only on $H$.
The proof above applies to this activity with adjusted constants.

Run the ideal process with failure probability $\delta/2$
for $M$ steps. Approximate each deletion probability
$\widehat\lambda^{-w_v/\overline w}$ by a dyadic rational
with absolute error at most $\delta/(2M)$.
These approximations can be computed and sampled using fair
random bits in time polynomial in the input length and
$\log(M/\delta)$.
Coupling the ideal and implemented processes bounds the
probability that their trajectories differ by $\delta/2$.
Thus the implementation succeeds with probability at least
$1-\delta$ and retains the EPTAS running time.
\end{remark}

\subsection{Arbitrary weights}

Consider Maximum Weight Independent Set on a star:
one center is adjacent to each of $m$ leaves, and
there are no edges between leaves. Give every leaf weight one and
the center weight $w_c>m$. The optimal solution is the center alone,
while every other solution has weight at most $m$.

Before the center is first selected, every proposed absent leaf
is accepted. Inserting the center requires a visit to the empty set.
The following theorem bounds the probability of inserting the center,
even when the activity is chosen adaptively.

\begin{theorem}\label{thm:star}
Let a star have $m\ge2$ leaves of weight one and a center of
weight $w_c>m$. Start weighted single-vertex Metropolis on
independent sets from the empty set, with independent uniform
vertex proposals. At each step $s$, the activity $\lambda_s\ge1$
may be chosen by an arbitrary randomized rule depending on the
entire past and the proposed vertex.
Let $\tau_c$ be the first time the center belongs to the current set.
Then, for every integer $t\ge1$,
\[
\Prb\{\tau_c\le t\}\le\frac{4+t2^{-m}}{m+1}.
\]
\end{theorem}

\begin{proof}
Consider a modified Metropolis process that stops as soon as
the center is inserted. It agrees with the original process
up to that moment, so the distribution of $\tau_c$ is unchanged.

For $0\le k\le m$, let $N_k$ be the expected number of indices
$s$ with $0\le s<\min\{t,\tau_c\}$ at which exactly $k$ leaves
are selected.
Then
\[
\sum_{k=0}^m N_k=\E[\min\{t,\tau_c\}]\le t.
\]

Whenever the selected set is empty, the next proposal
inserts the center with probability $1/(m+1)$.
Since this insertion ends the process,
\[
\Prb\{\tau_c\le t\}=\frac{N_0}{m+1}.
\]
It therefore suffices to bound $N_0$.

Fix $0\le k<m$. Between any two transitions from $k$ to $k+1$ selected leaves, the process must make a reverse transition from $k+1$ to $k$. Thus, if there are $q$ transitions from $k$ to $k+1$, there are at least $q-1$ reverse transitions.
Every absent leaf is added when proposed, whereas a proposed
deletion is accepted only with some probability.
Taking expectations gives
\[
\frac{m-k}{m+1}N_k
\le \frac{k+1}{m+1}N_{k+1}+1.
\]
These estimates hold for arbitrary activity choices based on
the history and the proposed vertex.

Using the identities for consecutive binomial coefficients,
we can rewrite this inequality as
\[
\frac{N_k}{\binom{m}{k}}
\le
\frac{N_{k+1}}{\binom{m}{k+1}}
+\frac{m+1}{m\binom{m-1}{k}}.
\]
Iterating from zero to $k$ yields
\[
N_0
\le
\frac{N_k}{\binom{m}{k}}
+\frac{m+1}{m}
 \sum_{j=0}^{k-1}\frac{1}{\binom{m-1}{j}}.
\]
Since $\binom{m-1}{j}\ge m-1$ for $1\le j\le m-2$, the second term can be bounded as follows:
\[
\frac{m+1}{m}
\sum_{j=0}^{m-1}\frac{1}{\binom{m-1}{j}}
\le
\frac{m+1}{m}
\left(2+\frac{m-2}{m-1}\right)
\le4.
\]
Thus
\[
N_0
\le
\frac{N_k}{\binom{m}{k}}+4.
\]

Multiplying this inequality by $\binom{m}{k}$ and summing
over $k=0,\ldots,m$, we obtain
\[
2^m(N_0-4)\le\sum_{k=0}^m N_k\le t.
\]
Consequently,
\[
\Prb\{\tau_c\le t\}
=\frac{N_0}{m+1}
\le\frac{4+t2^{-m}}{m+1}.
\]
\end{proof}

To see the approximation consequence, take $w_c=m^2$.
The optimum has weight $m^2$, whereas every set without the center
has approximation ratio at most $1/m$. For any fixed $\rho>0$,
a $\rho$-approximation therefore requires inserting the center when
$m>1/\rho$. With any polynomial number of proposals, the theorem
bounds the probability of this event by $O(1/m)$.
If $R$ is the approximation ratio of the best visited set, then
\[
\E R\le\Prb\{\tau_c\le t\}+1/m=O(1/m)
\]
for such a budget. Thus one trajectory from the empty set has no
uniform constant-factor guarantee for unrestricted weighted inputs,
even on planar trees.

Restarts change this behavior: each fresh run from the empty set
selects the center on its first proposal with probability $1/(m+1)$.

\section{Discussion}\label{sec:discussion}

The general theorem uses a division with few boundary vertices, many
useful replacements on that division, and a bound on the temporary
loss of each single-vertex conversion. The closure assumptions allow
all the selected separator vertices to be removed and restored as
needed. This gives the exponent $O_H(\eps^{-2})$ uniformly across
the properties considered here. The role of the fixed exterior is
explicit in the conversion lemma, which also applies to weighted
objectives.

The proof controls the time to reach a near-optimal state. It counts
improving paths from poor states and applies the resulting inequality
to trajectories that have not yet succeeded.
The factor $2^r$ comes from counting possible starting subsets inside
a region. A smaller count, or a different choice of improvements,
could improve the dependence on accuracy. Reducing the quadratic
dependence on $n$ is another open direction.

For weighted objectives, bounded weight imbalance preserves the
EPTAS guarantee. The weighted star gives an obstruction to polynomial-time
approximation for arbitrary weights in a single trajectory from the
empty set, even with adaptively chosen activities.

\section*{Use of AI tools}
The authors used ChatGPT (OpenAI) to assist in developing and refining the mathematical proofs, as well as for language editing. The authors reviewed the resulting material and take full responsibility for the content of the paper.


\begin{thebibliography}{99}

\bibitem{AST}
N. Alon, P. Seymour, and R. Thomas.
A separator theorem for nonplanar graphs.
\emph{Journal of the American Mathematical Society},
3(4):801--808, 1990.
\url{https://doi.org/10.1090/S0894-0347-1990-1065053-0}.

\bibitem{Baker}
B. S. Baker.
Approximation algorithms for NP-complete problems on planar graphs.
\emph{Journal of the ACM}, 41(1):153--180, 1994.
\url{https://doi.org/10.1145/174644.174650}.

\bibitem{BJM}
M. de Berg, B. M. P. Jansen, and D. Mukherjee.
Independent-set reconfiguration thresholds of hereditary graph classes.
\emph{Discrete Applied Mathematics}, 250:165--182, 2018.
\url{https://doi.org/10.1016/j.dam.2018.05.029}.

\bibitem{CG}
S. Cabello and D. Gajser.
Simple PTAS's for families of graphs excluding a minor.
\emph{Discrete Applied Mathematics}, 189:41--48, 2015.
\url{https://doi.org/10.1016/j.dam.2015.03.004}.

\bibitem{CMRW}
Z. Chen, D. Mikulincer, D. Reichman, and A. S. Wein.
Time lower bounds for the Metropolis process and simulated annealing.
In \emph{Approximation, Randomization, and Combinatorial Optimization.
Algorithms and Techniques (APPROX/RANDOM)}, volume 353 of \emph{LIPIcs},
pages 47:1--47:22, 2025.
\url{https://doi.org/10.4230/LIPIcs.APPROX/RANDOM.2025.47}.

\bibitem{CMZ}
Z. Chen, E. Mossel, and I. Zadik.
Almost-linear planted cliques elude the Metropolis process.
\emph{Random Structures \& Algorithms}, 66(2):e21274, 2025.
\url{https://doi.org/10.1002/rsa.21274}.

\bibitem{DT}
E. Davies and B. Tan.
Large independent sets via neighborhood block dynamics, Glauber dynamics,
and local occupancy.
arXiv:2611.02483, 2026.
\url{https://arxiv.org/abs/2611.02483}.

\bibitem{DHK}
E. D. Demaine, M. Hajiaghayi, and K. Kawarabayashi.
Algorithmic graph minor theory: Decomposition, approximation, and coloring.
In \emph{46th Annual IEEE Symposium on Foundations of Computer Science
(FOCS)}, pages 637--646, 2005.
\url{https://doi.org/10.1109/SFCS.2005.14}.

\bibitem{DR}
M. P. Desai and V. B. Rao.
Finite-time behavior of slowly cooled annealing chains.
\emph{Probability in the Engineering and Informational Sciences},
11(2):137--176, 1997.
\url{https://doi.org/10.1017/S0269964800004733}.

\bibitem{DRW}
B. Doerr, A. Rajabi, and C. Witt.
Simulated annealing is a polynomial-time approximation scheme
for the minimum spanning tree problem.
\emph{Algorithmica}, 86:64--89, 2024.
\url{https://doi.org/10.1007/s00453-023-01135-x}.

\bibitem{DSC}
P. Diaconis and L. Saloff-Coste.
Comparison theorems for reversible Markov chains.
\emph{Annals of Applied Probability}, 3(3):696--730, 1993.
\url{https://doi.org/10.1214/aoap/1177005359}.

\bibitem{DS}
P. Diaconis and D. Stroock.
Geometric bounds for eigenvalues of Markov chains.
\emph{Annals of Applied Probability}, 1(1):36--61, 1991.
\url{https://doi.org/10.1214/aoap/1177005980}.

\bibitem{EF}
D. Eppstein and D. Frishberg.
Rapid mixing for the hardcore Glauber dynamics and other Markov chains
in bounded-treewidth graphs.
In \emph{34th International Symposium on Algorithms and Computation
(ISAAC)}, volume 283 of \emph{LIPIcs}, pages 30:1--30:13, 2023.
\url{https://doi.org/10.4230/LIPIcs.ISAAC.2023.30}.

\bibitem{Grohe}
M. Grohe.
Local tree-width, excluded minors, and approximation algorithms.
\emph{Combinatorica}, 23(4):613--632, 2003.
\url{https://doi.org/10.1007/s00493-003-0037-9}.

\bibitem{Hajek}
B. Hajek.
Cooling schedules for optimal annealing.
\emph{Mathematics of Operations Research}, 13(2):311--329, 1988.
\url{https://doi.org/10.1287/moor.13.2.311}.

\bibitem{HOST}
H. P. Hoang, N. Ohsaka, R. Saito, and Y. Tamura.
On (in)approximability of MaxMin independent set reconfiguration.
In \emph{53rd International Colloquium on Automata, Languages, and
Programming (ICALP)}, volume 374 of \emph{LIPIcs},
pages 108:1--108:16, 2026.
\url{https://doi.org/10.4230/LIPIcs.ICALP.2026.108}.

\bibitem{Jerrum}
M. Jerrum.
Large cliques elude the Metropolis process.
\emph{Random Structures \& Algorithms}, 3(4):347--359, 1992.
\url{https://doi.org/10.1002/rsa.3240030402}.

\bibitem{KGV}
S. Kirkpatrick, C. D. Gelatt, Jr., and M. P. Vecchi.
Optimization by simulated annealing.
\emph{Science}, 220(4598):671--680, 1983.
\url{https://doi.org/10.1126/science.220.4598.671}.

\bibitem{LT}
R. J. Lipton and R. E. Tarjan.
Applications of a planar separator theorem.
\emph{SIAM Journal on Computing}, 9(3):615--627, 1980.
\url{https://doi.org/10.1137/0209046}.

\bibitem{LMRRW}
K. Liu, S. Mohanty, P. Raghavendra, A. Rajaraman, and D. X. Wu.
Locally stationary distributions: A framework for analyzing slow-mixing
Markov chains.
In \emph{65th IEEE Annual Symposium on Foundations of Computer Science
(FOCS)}, pages 203--215, 2024.
\url{https://doi.org/10.1109/FOCS61266.2024.00022}.
Full version: \url{https://arxiv.org/abs/2405.20849}.

\bibitem{MRRTT}
N. Metropolis, A. W. Rosenbluth, M. N. Rosenbluth,
A. H. Teller, and E. Teller.
Equation of state calculations by fast computing machines.
\emph{Journal of Chemical Physics}, 21(6):1087--1092, 1953.
\url{https://doi.org/10.1063/1.1699114}.

\bibitem{SH}
G. H. Sasaki and B. Hajek.
The time complexity of maximum matching by simulated annealing.
\emph{Journal of the ACM}, 35(2):387--403, 1988.
\url{https://doi.org/10.1145/42282.46160}.

\bibitem{Sinclair}
A. Sinclair.
Improved bounds for mixing rates of Markov chains and multicommodity flow.
\emph{Combinatorics, Probability and Computing}, 1(4):351--370, 1992.
\url{https://doi.org/10.1017/S0963548300000390}.

\bibitem{Tazari}
S. Tazari.
Faster approximation schemes and parameterized algorithms on
$H$-minor-free and odd-minor-free graphs.
In \emph{Mathematical Foundations of Computer Science (MFCS)},
volume 6281 of \emph{Lecture Notes in Computer Science},
pages 641--652, 2010.
\url{https://doi.org/10.1007/978-3-642-15155-2_56}.

\bibitem{Thomason}
A. Thomason.
The extremal function for complete minors.
\emph{Journal of Combinatorial Theory, Series B}, 81(2):318--338, 2001.
\url{https://doi.org/10.1006/jctb.2000.2013}.

\bibitem{Wild}
D. S. Wild, D. Sels, H. Pichler, C. Zanoci, and M. D. Lukin.
Quantum sampling algorithms, phase transitions, and computational complexity.
\emph{Physical Review A}, 104:032602, 2021.
\url{https://doi.org/10.1103/PhysRevA.104.032602}.

\end{thebibliography}
\end{document}